\documentclass{article}
\usepackage[a4paper, left=1.5cm, right=1.5cm, top=2.5cm, bottom=2.5cm]{geometry}
\usepackage{graphicx} 
\usepackage{amsmath}
\usepackage{amssymb}
\usepackage{authblk}
\usepackage{cite}
\title{Thermodynamic Split Conjecture and an Observational Test for Cosmological Entropy}
\author{Oem Trivedi \thanks{oem.trivedi@vanderbilt.edu}}
\affil{Department of Physics and Astronomy, Vanderbilt University, Nashville, TN 37235, USA}
\date{\today}

\begin{document}

\maketitle

\begin{abstract}
We revisit string theoretic derivations of black hole entropy and argue that their enabling structures do not persist in realistic cosmologies. We formalize this as the \textit{Thermodynamic Split Conjecture} (TSC) which is the statement that in any UV complete quantum gravity, black hole and cosmological horizon thermodynamics are generically inequivalent. The BKE criterion is then formulated to formalize this approach while we also discuss ways to falsify the conjecture. Finally, we propose an observational scaling test which is centered around comparisons of data native entropy proxies from tomographic maps to the Bekenstein–Hawking prediction \(S\propto H^{-2}(z)\). The framework both sharpens when and why the area law holds and provides a roadmap towards a UV complete description of cosmological horizon entropy.
\end{abstract}

\section{Introduction}

Cosmology is possibly going through a golden age right now, with theory and observation advancing in quite an unprecedented fashion to probe the Universe across incredible ranges of scale and precision. The discovery and modeling of dark energy \cite{de1SupernovaSearchTeam:1998fmf,de2Li:2012dt,de3Li:2011sd,de4Mortonson:2013zfa,de5Frusciante:2019xia,de6Huterer:2017buf,de7Vagnozzi:2021quy,de8Adil:2023ara,de9Feleppa:2025clx,de10DiValentino:2020evt,de11Nojiri:2010wj,de12Nojiri:2006ri,de13Trivedi:2023zlf,de14Trivedi:2022svr,de15Trivedi:2024inb,de16Trivedi:2024dju} and dark matter \cite{dm11rubin1970rotation, dm1Cirelli:2024ssz,dm2Arbey:2021gdg,dm3Balazs:2024uyj,dm4Eberhardt:2025caq,dm5Bozorgnia:2024pwk,dm6Misiaszek:2023sxe,dm7OHare:2024nmr,dm8Adhikari:2022sbh,dm9Miller:2025yyx,dm10Trivedi:2025vry} have opened rich theoretical landscapes, while persistent anomalies and cosmic tensions like the $H_0$ and $S_8$ tensions \cite{ht1DiValentino:2021izs,ht2Clifton:2024mdy,s81kazantzidis2018evolution,s82amon2022non,s83poulin2023sigma,s84Ferreira:2025lrd,baoFerreira:2025lrd} continue to challenge our standard frameworks and motivate new ideas. At the same time, a new observational era is coming online, spanning next-generation CMB experiments, galaxy and weak-lensing surveys, multi-messenger astrophysics,and precision probes of the early Universe \cite{t1jha2019next,t2Shanks:2015lda,t3chandler2025nsf,t4Euclid:2024yrr,t5WST:2024rai,t6CosmoVerseNetwork:2025alb,t7COMPACT:2022gbl}. All of this sharpen the theoretical questions one can possibly think about with regards to gravity and spacetime on cosmological scales.
\\
\\
Perhaps in order to understand some of the most fundamental questions in cosmology, one has to turn their attention towards a theory of quantum gravity. In this regard string theory has arguably been the most famous of such paradigms so far. It offers extended objects (strings and branes), extra dimensions, dualities, and powerful holographic tools that provide nonperturbative control in certain regimes. Compactifications and fluxes yield a broad space of effective low-energy theories, and gauge/gravity correspondences relate strongly coupled quantum systems to semiclassical gravities \cite{st1Green:1984sg,st2Schwarz:1982jn,st3Veneziano:1968yb,st4Banks:1996vh,st5Sen:2024nfd,st6Cicoli:2023opf,st7Nielsen:1973cs}. One promising direction that has been explored using string theory is the resolution of the anomalous properties of the black hole, in particular what happens inside the horizon, the information paradox and the entropy.
\\
\\
This line of inquiry has thrown up quite a few interesting proposals from string theory aimed at addressing them and some of these are very non-trivial as well. These approaches includes Fuzzballs based on microstate geometries, D-brane microstate counting methods (Strominger–Vafa and extensions), AdS/CFT state counting methods with Cardy-type arguments, the Quantum Entropy Function(QEF) approach, various attractor mechanisms with higher-derivative Wald entropy, the Ooguri–Strominger–Vafa topological string approach, Horowitz and Polchinki's String–black hole correspondence and recently the Islands program. Over the past two decades, these frameworks have in their different capacities explained key aspects of black hole thermodynamics in controlled settings and has been able to reproduce entropy via microstate counting in different ways, and perhaps even offering a unitary picture for radiation and information recovery. Semiclassically, however, the same Bekenstein-Hawking entropy formula \cite{sb1hawking1972black,sb2bekenstein1973black,sb3bekenstein1974generalized,sb4Cotler:2016fpe,sb5krolak1978singularities,sb6clarke1985conditions,sb10penrose1965gravitational,sb9senovilla20151965} is often applied to cosmological horizons (like the dS/Gibbons-Hawking case). This is done very freely in theoretical cosmology and has led to a huge body of literature as well. But one can take a step back and ask a critical question; does a UV-complete theory of gravity allow for using the black hole formulations to the cosmological event horizon? In this work we examine this question in detail and we structure it as follows. In Section II we give a concise overview of the various proposals in String theory aimed at addressing black hole entropy while in Section III we discuss the main conceptual difficulties that arise when attempting to construct a cosmological analogue and in Section IV we introduce the Thermodynamic Split Conjecture and discuss a mathematical form for it in detail. We discuss the implications of this conjecture on cosmology in section V, propose an observational test for Bekenstein-Hawking entropy for an expanding cosmology in section VI and we finally conclude in Section VII with a summary and outlook.
\\
\\
\section{Black Hole Entropy formulations in String Theory}
We begin with the microstate counting procedures which have been explored to study black entropies. In the original D-brane approach, one makes BPS black holes as bound states of branes whose low-energy worldvolume dynamics is a supersymmetric quantum field theory or a two-dimensional CFT, with Strominger and Vafa providing a way forward in this regard initially \cite{sv1Strominger:1996sh,sv2Nishioka:2009un,sv3Strominger:1997eq,sv4Bena:2005va,sv5Carlip:1998wz,sv6DeHaro:2019gno,sv7Emparan:2006it,sv8Mayerson:2020acj,sv9Sen:1999mg}. The key step is to count protected degeneracies in a fixed charge sector \((\{Q_i\})\) and with the protection ensuring invariance as one moves from weak coupling( where the counting is tractable) to strong coupling( where the gravitational black hole exists). In the D1–D5–P system, for example, the relevant CFT has central charge \(c=6n_1 n_5\) and the asymptotic degeneracy of left/right-moving excitations at levels \((N_L,N_R)\) is controlled by modular invariance via Cardy-asymptotics \cite{car1Cardy:1986ie,car2Cardy:1989ir,car3Averin:2019zsi} and this is the key in reproducing the macroscopic Bekenstein–Hawking entropy. In its canonical Cardy form this is
\begin{equation}
S_{\text{CFT}} \;=\; 2\pi\!\left(\sqrt{\frac{c\,L_0}{6}}+\sqrt{\frac{c\,\bar L_0}{6}}\right)
\end{equation}
upon identifying the CFT quantum numbers with the brane charges, this matches to
\begin{equation} \label{sbh}
S_{\text{BH}} \;=\; \frac{A_{\text{hor}}}{4G_N}
\end{equation}
This logic extends far beyond the original five-dimensional example to many four-dimensional \(\mathcal{N}{=}2\) setups as well which is where the microscopic degeneracies are packaged by automorphic forms and the large-charge growth precisely reproduces the area law and this also leads in cases including controlled subleading corrections. Closely linked to this is the AdS/CFT route \cite{adscftMaldacena:1997re}, where one does not directly count brane bound states at weak coupling but instead uses the dual CFT at strong coupling. The paradigmatic case is the BTZ black hole in AdS\(_3\) with Brown–Henneaux central charge \(c=\tfrac{3\ell}{2G_3}\). In this case the modular invariance and the Cardy formula then give
\begin{equation}
S_{\text{BTZ}} \;=\; \frac{2\pi r_+}{4G_3}
\end{equation}
which equals the Bekenstein–Hawking entropy as well \eqref{sbh}. The similarity between the two paths is the reliance on a well-defined boundary CFT with modular properties and the difference is methodological, with the D-brane derivation using indices in a weakly coupled description, and AdS/CFT using thermodynamics of the strongly coupled dual.
\\
\\
Another arena where microstates have a central importance is fuzzball proposal of Mathur, which is formulated primarily within the framework of type IIB string theory and seeks to resolve the information paradox \cite{ip1Hawking:2015qqa,ip2Raju:2020smc,ip3Polchinski:2016hrw} by positing that what we perceive as a black hole is actually an ensemble of horizonless, smooth, and quantum gravitational configurations \cite{fuz1Mathur:2005zp,fuz2Mathur:2009hf,fuz3Freedman:1998tz,fuz4Lunin:2001jy,fuz5Das:1996wn,fuz6Das:1996we,fuz7Brower:2000rp,fuz8Mathur:2008nj,fuz9Mathur:2005ai}. These geometries differ from the classical black hole at horizon scale and are often composed of bound states of strings and branes, such as D1-D5-P systems and so every fuzzball solution corresponds to a distinct microstate of the black hole and the statistical ensemble of these states reproduces the Bekenstein-Hawking entropy formula. In favorable cases this has been demonstrated explicitly, for instance in the D1-D5 system, where microstate counting using various string dualities and supersymmetry algebra leads to the entropy expression
\begin{equation}
S_{\text{BH}} = 2\pi \sqrt{n_1 n_5 n_p}
\end{equation}
which agrees with the Bekenstein-Hawking result for the corresponding black hole. Importantly we see that these geometries are smooth and horizonless which goes to say that they evade the problems associated with the event horizon and singularity, and so we can say that they work towards providing a unitarily evolving quantum description. It is important to note that the the success of the fuzzball program rests on several string theoretic ingredients and caveats. It needs the presence of extended objects like D-branes and fundamental strings which is essential for constructing microstates,  compactification of extra dimensions etc. While most fuzzball solutions are only applicable in asymptotically flat or AdS spacetimes
\\
\\
A complementary macroscopic route is provided by QEF methods championed by Sen, where one computes the exact degeneracy associated with the near-horizon AdS\(_2\) of extremal black holes through the means of a gravitational path integral with fixed electric/magnetic charges at the AdS\(_2\) boundary \cite{qe1Sen:2008vm,qe2Sen:2009vz,qe3Sen:2012kpz,qe4Banerjee:2010qc,qe5Banerjee:2011jp,qe6Sen:2012cj,qe7Dabholkar:2014ema,qe8Castro:2009jf,qe9Castro:2008ms,qe10Dabholkar:2010rm,qe11Benini:2017oxt,qe13Sen:2014aja,qe14Iliesiu:2022kny,qe15Iliesiu:2022onk,qe16Cabo-Bizet:2017jsl}. In this framework one evaluates
\begin{equation}
d(\{q,p\}) \;=\; Z_{\text{AdS}_2}[q,p] \;\sim\; \exp\!\left(S_{\text{Wald}}(\{q,p\}) + \text{quantum corrections}\right)
\end{equation}
so that the statistical entropy \(S_{\text{micro}}=\log d\) matches the macroscopic result including higher-derivative terms and loops. The QEF links deeply with the attractor \cite{att1Sen:2005wa,att2Sen:2007qy,att3Tripathy:2005qp,att4Alishahiha:2006ke,att5Cardoso:2006xz,att6Alishahiha:2006jd} mechanism of \(\mathcal{N}{\ge}2\) supergravity, wherein it is well known that the near-horizon values of moduli extremize an entropy or central-charge functional and it ends up fixing them purely in terms of the conserved charges. In a convenient formulation one usually works with Sen’s entropy function \(E\)
\begin{equation}
E(e_i,p^i,u^a) \;=\; 2\pi \!\left(q_i e^i - f(e_i,p^i,u^a)\right)
\end{equation}
where \(f\) is the Lagrangian density integrated over the horizon, \(e_i\) are near-horizon electric fields, \(p^i\) magnetic charges, and \(u^a\) moduli; extremization \(\partial E{= }0\) yields both the attractor equations and the entropy
\begin{equation}
S_{\text{macro}} \;=\; E\big|_{\text{ext}} \;=\; S_{\text{Wald}} \;+\; \cdots
\end{equation}
Here \(S_{\text{Wald}}\) is the Noether-charge entropy appropriate to higher-derivative gravity and
\begin{equation}
S_{\text{Wald}} \;=\; -2\pi \int_{\mathcal{H}}\! d^{d-2}\Sigma \;\frac{\partial \mathcal{L}}{\partial R_{abcd}}\,\epsilon_{ab}\epsilon_{cd}
\end{equation}
Note that the ellipsis denotes quantum and string based corrections that the QEF systematically incorporates and these macroscopic constructions provide a regulator independent target that the microscopic degeneracies hit. So, we see thereby that it ends up elevating the area law to an exact statement within its regime of applicability.
\\
\\
Another very powerful organizing principle relating the microscopic and macroscopic sides is the Ooguri–Strominger–Vafa proposal. The heart of this is the idea that one expresses the mixed ensemble black hole partition function in terms of the topological string partition function on the Calabi–Yau target of the compactification \cite{osv1Ooguri:2004zv,osv2Denef:2007vg,osv3Bena:2007kg,osv4Kraus:2005vz,osv5Sen:2005iz,osv6Sen:2012kpz,osv7LopesCardoso:2004law,osv8Gupta:2012cy,osv9Castro:2007ci,osv10Cribiori:2022nke,osv11Dedushenko:2014nya,osv12Saraikin:2007jc,osv13Cribiori:2023ffn}. We start here by denoting collectively the magnetic charges by \(p\) and the electric chemical potentials by \(\phi\) and then write schematically
\begin{equation}
Z_{\text{BH}}(p,\phi) \;\approx\; \big|Z_{\text{top}}(g_s,t)\big|^2
\end{equation}
with the map between \((p,\phi)\) and topological data \((g_s,t)\) determined by special geometry. There is a Legendre transform in \(\phi\) which then reproduces \(S_{\text{BH}}\) in a large-charge expansion and what is interesting is that the higher-genus topological string amplitudes encode the infinite tower of subleading corrections to the area law. Although this is indeed very conjectural and sensitive to measure choices, this bridge has yielded many nontrivial checks of microscopic formulas against macroscopic entropies in \(\mathcal{N}{=}2\) compactifications.
\\
\\
From a different angle, the string–black hole correspondence of Horowitz and Polchinski \cite{hp1Horowitz:1996nw,hp2Witten:1998zw,hp3Ashtekar:2000eq,hp4Mohaupt:2000mj,hp5Horowitz:1997jc,hp6Sfetsos:1997xs,hp7Solodukhin:1997yy,hp8Visser:1997yu,hp9Balasubramanian:2007bs,hp10Stoffers:2012mn,hp11Silverstein:2003jp,hp12Aalsma:2019ryi,hp13Sheinblatt:1997nt} provides a continuity argument rather than a precise degeneracy calculation. Highly excited single long strings of mass \(M\) here have a Hagedorn growth of states \(S_{\text{string}}\!\sim\! 2\pi \sqrt{N}\) with level \(N\!\sim\!\alpha' M^2\) and as the string coupling or effective size grows, there is a correspondence point where the self-gravitating string’s Schwarzschild radius \(r_s\) becomes of order the string length \(\ell_s\), and the state crosses over into a black hole. At this point the entropies match parametrically and lead to 
\begin{equation}
S_{\text{string}}(N) \;\sim\; S_{\text{BH}}(M) \;\sim\; \frac{A(r_s\!\sim\!\ell_s)}{4G_N}
\end{equation}
This hence gives us an explanation for the area law’s emergence from string microphysics and for why the microscopic and macroscopic descriptions must agree in their overlapping domain. While not a substitute for exact counting, one can definitely see that this correspondence underpins the expectation that the area law is the universal coarse-grained limit of an underlying stringy spectrum.
\\
\\
Finally, recent progress on the black hole information problems within semiclassical gravity has led to the use of the island prescription to compute the fine-grained entropy of Hawking radiation through a generalized entropy functional. The core idea is that given a candidate “island” region \(I\) behind the horizon \cite{i1Geng:2020qvw,i2Hashimoto:2020cas,i3Geng:2021hlu,i4Krishnan:2020fer,i5Bhattacharya:2021jrn,i6Karch:2023ekf}, one extremizes
\begin{equation}
S_{\text{gen}}[I] \;=\; \frac{\mathrm{Area}(\partial I)}{4G_N} \;+\; S_{\text{out}}(\Sigma\setminus I)
\end{equation}
and then goes on to select the minimal quantum extremal surface. This procedure goes to reproduce the Page curve in holographic model where at early times the minimal configuration has no island and one recovers the familiar thermal growth while after the Page time, an island appears and the area term \(\mathrm{Area}(\partial I)/(4G_N)\) caps the entropy which thus leads to the restoration of unitarity as well. Although this is not a microscopic brane/CFT count, the appearance and dominance of the area term in \(S_{\text{gen}}\) explains \footnote{Note that this is also only true  within a controlled gravitational path integral.} why the Bekenstein–Hawking contribution governs the leading thermodynamics of horizons and how quantum corrections reorganize the entropy in time-dependent evaporation.
\\
\\
So to give a general comment, one has microscopic methods based on D-branes, AdS/CFT count states directly and Fuzzballs which can reproduce \(S_{\text{BH}}\) with precision, while macroscopic methods based on QEF, attractors, and Wald entropy compute the same quantity from the gravitational side including higher-derivative and quantum effects. Then we have  OSV which relates both to topological amplitudes that generate the subleading tower, while the Horowitz-Polchinski correspondence and islands frameworks explain the parametric continuity with string spectra and the dynamical role of the area term in fine-grained entropy, respectively. Together, we see that string theory offers a vast array of ways to reproduce the Bekenstein-Hawking formulation.
\\
\\
\section{Issues Extending it to Cosmology}
Now we shall discuss why there are issue extending the formulations of the approaches discussed above to the case of cosmological event horizons instead of black hole event horizons, for which they were constructed. This is meant to tackle the question of whether, given an underlying quantum gravity theory, we are allowed to just transplant the Black hole thermodynamic formulas to the case of a cosmology, which is something quite ubiquitous in the literature for the past three decades.
\\
\subsection*{No Holographic Boundary and Loss of Modular Control}
All successful string theoretic entropy computations lean on some well posed notions of asymptotics that supports a dual quantum system with powerful thermodynamic structure and so holography is at the heart of things \cite{hol1Susskind:1994vu,hol2tHooft:1993dmi,hol3tHooft:1990fkf,hol4Bousso:2002ju,hol5Susskind:1998dq,hol6Fischler:1998st,hol7Bianchi:2001kw}. The lack of a holographic framework in dS space is a significant issue as unlike AdS/CFT, no similarly concrete framework exists for dS/CFT. While indeed there are proposals like the Strominger dS/CFT correspondence \cite{dscftStrominger:2001pn} \footnote{There is even a Kerr/CFT correspondence \cite{kerrcftGuica:2008mu}, which the reader might be interested in.}, they are quite speculative and not equipped with nearly as much computational machinery or symmetry constraints as the AdS/CFT counterpart(which has made it successful).
\\
In AdS/CFT, a timelike conformal boundary hosts the CFT and modular invariance underwrites Cardy growth. One then computes microscopic degeneracies that match the area law. On the other hand in the D-brane approach, even when phrased without explicit AdS language, one effectively exploits a decoupling limit that yields a controllable boundary CFT or worldvolume theory. Cosmological spacetimes such as FLRW or global dS admit no comparably predictive boundary as there is no timelike infinity on which to anchor a dual QFT, nor a modular bootstrap that fixes the asymptotic density of states. Without a holographic boundary, both the CFT state-counting route and the gravitational island constructions lose their principal organizing principle. So, the very object whose degeneracy should reproduce \(A/4G\) is no longer defined in an observer independent manner.
\subsection*{Absence of Asymptotic Charges and Superselection Sectors}
Microscopic counting in string theory organizes the Hilbert space into so-called superselection sectors, which are labeled by conserved charges defined at infinity and these include D-brane numbers, electric/magnetic charges, angular momenta and the ADM mass. These are extracted from Gauss law type integrals on a sphere at infinity and produce the fixed-\(\{Q_i\}\) microcanonical ensemble whose degeneracy \(\Omega(E,\{Q_i\})\) one compares to \eqref{sbh}. Cosmological geometries provide no such \(S^{d-2}_\infty\) though, and observers are confined to finite causal patches with no gauge-invariant way to assign global \(\{Q_i\}\). This erases the primary tool used in D-brane counts, thus obstructs the charge-fixing that underlies the attractor mechanism and deprives the OSV framework of its canonical mixed ensemble. Without charges at infinity, there is no unambiguous sector in which to define a microscopic degeneracy that could match a cosmological horizon area.
\\
\subsection*{No Global Time-Translation Symmetry and the Failure of Gibbs/KMS approach}
Black hole thermodynamics rests on a Killing horizon with constant surface gravity and a global KMS state with respect to a timelike Killing vector. This is from where the zeroth and first laws, the Tolman redshifts, and the Euclidean periodicity follow. Expanding cosmologies are genuinely time dependent and admit no global timelike Killing field as for example, even in dS the thermal description is confined to a static patch and is explicitly observer-dependent. As a consequence, there is no global Gibbs ensemble or canonical temperature with which to formulate a universal horizon thermodynamics. Cardy-type arguments presuppose equilibrium thermodynamics in the dual CFT \cite{car1Cardy:1986ie,car2Cardy:1989ir}, while the island replica method \cite{i1Geng:2020qvw,i4Krishnan:2020fer,i2Hashimoto:2020cas} presumes a stationary setup or a bath with controlled time evolution. It is then important to note that in cosmology, these assumptions fail at the most basic level as the non equilibrium character of FLRW thus blocks the very thermodynamic ladder by which one ascends from microstates to area-law entropy.
\subsection*{No Universal Near-Horizon Throat and Attractor Control}
On the macroscopic side, exact entropy matches rely on a universal near-horizon structure that isolates the relevant degrees of freedom and renders the gravitational path integral well defined. Extremal black holes exhibit AdS\(_2\) throats and attractor flows that fix moduli purely in terms of charges. Wald’s Noether-charge entropy refined by the Quantum Entropy Function \cite{qe1Sen:2008vm,qe2Sen:2009vz}, then reproduces the microscopic degeneracy including higher-derivative and quantum corrections. Cosmological horizons lack these features as there is no AdS\(_2\) region, no attractor extremization, and no accepted boundary conditions that would make a QEF-like computation unambiguous. Even in the de Sitter static patch the absence of asymptotic charges and the observer dependence of the patch frustrate any attempt to transplant the extremization machinery. So we see that the macroscopic target that is so sharply defined in black holes becomes regulator and slicing dependent in cosmology, depriving microscopic approaches of a trustworthy quantity to match in this case and to produce the necessary results. 
\subsection*{Supersymmetry Breaking and the Loss of Index Protection}
The precision of string theoretic entropy counts is inseparable from supersymmetry as BPS bounds, protected indices, and non-renormalization theorems ensure that degeneracies computed at weak coupling survive to the strong-coupling black hole regime. Realistic cosmologies break supersymmetry at a high scale and evolve through time-dependent moduli spaces and so without unbroken supersymmetry, indices cease to protect degeneracies. This leads to the wall crossing becoming time dependent rather than parametric and so the worldsheet  control that underlies both D-brane counting and OSV is destructed. The very reason microscopic formulas hit the macroscopic target in black holes is SUSY and as that has no analogue in an expanding Universe, attempts to extrapolate those formulas to cosmology trade precision for speculation.
\subsection*{Observer Dependence, Entanglement Anchoring, and the Area Term}
In AdS/CFT, Ryu–Takayanagi \cite{Ryu:2006bv} and its covariant generalizations anchor extremal surfaces to a boundary and so to compute entanglement entropies that match horizon areas in the appropriate limits. In cosmology there is no boundary that one can anchor to and so any entanglement notion is inherently relational as one computes entropy across causal diamonds or observer dependent patches. The island program does reproduce area terms and Page curve methods in carefully engineered holographic models, but its cosmological generalization is ambiguous because one lacks a clean bath, a factorized Hilbert space, and a unique choice of replica contour. Even granting a dS static patch with a Gibbons–Hawking temperature, the resulting thermal nature is just tied to a particular observer and does not define an observer-independent microstate degeneracy. So it means that while area functionals remain useful in semiclassical gravity, their microscopic meaning in cosmology is different in kind from the charge-labeled degeneracies of black holes.
\\
\subsection*{BPS, AdS and Structure Reliance}
Many of the most precise entropy derivations in string theory rest on special structural pillars that are absent in cosmology. On the microscopic side, D–brane counts, protected indices, the OSV relation and the quantum entropy function are typically developed for BPS or near–BPS configurations in which  the BPS bound \cite{bogomol1976stability,psprasad1975exact} is saturated,
\begin{equation}
M \;=\; |Z(\{Q_i\})|
\end{equation}
and non–renormalization theorems ensure that degeneracies computed at weak coupling match strong–coupling black holes. This BPS structure also underlies attractor flows and the existence of a universal near–horizon geometry that cleanly isolates the degrees of freedom contributing to the entropy. Away from BPS control, it so happens that indices lose protection, wall–crossing becomes dynamical rather than parametric and neither microstate degeneracies nor macroscopic extremization principles remain sharp. Cosmologies are generically non–supersymmetric and time–dependent, so the very structure that makes the black hole area law computable is missing from the outset.
\\
Equally important is the reliance on AdS structure as AdS/CFT state counting presupposes a timelike conformal boundary and modular properties of the dual CFT. QEF computations exploit AdS\(_2\) throats with well–posed boundary conditions and even island calculations achieving unitary Page curves are most controlled in AdS–like, bath–coupled setups. Cosmological spacetimes offer neither a timelike boundary nor a universal AdS\(_2\) near–horizon region, and any static–patch thermality in de Sitter is explicitly observer–dependent. This AdS dependence is therefore a genuine obstruction to transplanting the black hole entropy machinery to FLRW/dS backgrounds.
\\
\subsection*{Limits of the Vecro-based Cosmological Fuzzball Proposal}
The vecro picture \cite{elasMathur:2021zzr} treats the quantum vacuum as populated by rare, highly degenerate virtual fluctuations of horizon-sized black hole microstates or vecros. In gravitational collapse, compression across a trapped surface is argued to nucleate on-shell fuzzballs and so in an expanding universe by contrast, the stretching of vecros across anti-trapped surfaces during changes of the expansion law is posited to inject an effective vacuum energy. A convenient phenomenological parameterization is to encode the injection at a transition time \(t_t\) as
\begin{equation}
\Delta \rho_{\rm vac}(t_t)\;\sim\;\alpha\, M_{\rm Pl}^2 H^2(t_t)
\end{equation}
with a small, dimensionless amplitude \(\alpha\) controlled heuristically by the adiabaticity of the transition. This yields an EDE-like bump near equality if \(\alpha>0\), while giving negligible effect in exact radiation domination. As a qualitative selection rule this is appealing and consistent with basic cosmological constraints.
\\
However, the proposal does not supply the structural ingredients that made black hole entropy calculations precise and it is because of multifold reasons. First, there is no holographic dual or boundary theory from which to count vecro states and so the degeneracy that supposedly compensates action suppression is asserted rather than derived. Without asymptotic charges there is no observer independent superselection sector \(\{Q_i\}\) in which to define a microcanonical degeneracy \(\Omega(E,\{Q_i\})\), so the central identity \(S_{\rm micro}=\log\Omega\simeq A/4G\) has no cosmological analogue in this framework. Secondly the expansion history breaks global time-translation symmetry and precludes a Killing horizon with a global KMS state and so the vecro mechanism therefore does not reinstate equilibrium thermodynamics or a canonical temperature beyond an observer’s static patch. Also, the macroscopic side lacks a universal near horizon throat that would render an entropy functional regulator independent. In particular there is no demonstrated gravitational path integral that isolates a cosmological “microstate entropy’’ tied to an area term.
\\
Finally, to make this more concrete, one would hope that at the effective theory level the stress tensor of stretched vecros is not derived from a UV-complete action and its conservation is imposed rather than proven. One would like to see a covariant prescription for \(T^{\mu\nu}_{\rm vecro}\) such that
\begin{equation}
\nabla_\mu\!\left(T^{\mu\nu}_{\rm tot}\right)=\nabla_\mu\!\left(T^{\mu\nu}_{\rm std}+T^{\mu\nu}_{\rm vecro}\right)=0
\end{equation}
hold across transitions without violating causality or introducing superluminal sound speeds in perturbations. Without such a formulation, the sign, time profile, and clustering properties of \(\Delta \rho_{\rm vac}\) remain free parameters rather than predictions.Because the vecro sector is invoked precisely when anti-trapped surfaces grow, its contribution is intrinsically patch dependent too which again reinforces the observer relational character of cosmological “thermodynamics". So the vecro approach offers an interesting phenomenology for transient vacuum-energy injection and is compatible with swampland cautions about stable de Sitter vacua, but it does not resolve the core obstacles that obstruct a cosmological analogue of stringy black hole entropy. It provides neither a boundary counting nor an attractor/QEF macroscopic target and it does not restore the equilibrium structure needed for a Bekenstein–Hawking type microstate interpretation of cosmological horizons.
\\
\\
\section{Thermodynamic Split Conjecture and its Implications}
The preceding section established the crucial point that constructing analogous descriptions for cosmological horizons based on microstate-couting or other such descriptions as for the black hole horizons fail due to a combination of technical and conceptual obstructions rooted in the very nature of these approaches. This includes the lack of holographic duals, structure unavailability etc. and these differences suggest a profound possibility; Thermodynamics for a cosmology, especially in connection with cosmological horizons, might be fundamentally distinct from black hole thermodynamics and cannot be naively extrapolated from it.
\\
\\
Building from this point, we are led to formalize the separation between black hole and cosmological horizon thermodynamics as a precise conjecture about the microscopic and macroscopic structures that are at the heart of the Area law.

\medskip
\noindent\textit{Thermodynamic Split Conjecture (TSC):}
\textit{In any UV complete theory of quantum gravity, the microscopic and thermodynamic structures associated with black hole event horizons and with cosmological event horizons are generically inequivalent.}
\medskip

By the term “generically inequivalent’’, we mean that it asserts that there is no observer–independent identification between a black hole microcanonical ensemble labeled by conserved charges and any ensemble for a cosmological horizon that reproduces the same entropy functional. Consequently, the black hole area law arises from charge labeled microstate counting in controlled regimes, whereas the cosmological “area terms’’ belong, at best, to relational/entanglement constructs tied to causal patches and do not admit a universal microstate interpretation. Note again that here we assume that the implications drawn from string theory as in section 3 could hold true for the UV complete quanutm gravity theory. 
\\
\subsection*{BKE Criterion}
To make this precise, we encode the structural pillars of black hole entropy by a trio of conditions \((B,K,E)\). These all can take binary values which are \(\{0,1\}\). Consider now that \((\mathcal{M},g)\) is a globally hyperbolic spacetime with a future event horizon \(\mathcal{H}^+\subset\mathcal{M}\), and let \(\mathcal{Q}\) denote the set of conserved charges of the gravitational plus matter system. With this, we define the following:

\medskip
\emph{(B) Boundary/charges.} We set \(B=1\) if there exists a nontrivial asymptotic region \(\partial\mathcal{M}_\infty\) such that \(\mathcal{Q}\) contains a complete set of conserved charges \(\{Q_i\}\) defined by Gauss law fluxes
\begin{equation}
Q_i \;=\; \oint_{S^{d-2}_\infty}\!\!*J_i \,, \qquad \frac{dQ_i}{dt}=0
\end{equation}
which label superselection sectors of the Hilbert space \(\mathcal{H}=\bigoplus_{\{Q_i\}}\mathcal{H}_{\{Q_i\}}\). If this is not satisfied, then B is set to 0.

\medskip
\emph{(K) Killing/Gibbs.} We set \(K=1\) if there exists a globally defined timelike Killing vector \(\chi^a\) that becomes null on \(\mathcal{H}^+\) \footnote{This essentially means that there is a Killing horizon.}, with constant surface gravity \(\kappa\) and a KMS state for QFT on \((\mathcal{M},g)\) at inverse temperature \(\beta=2\pi/\kappa\),
\begin{equation}
\nabla_{(a}\chi_{b)}=0,\qquad \chi^2\big|_{\mathcal{H}^+}=0,\qquad \kappa=\text{const},\qquad \langle O(t+i\beta)O(0)\rangle=\langle O(t)O(0)\rangle
\end{equation}
This would ensure that Tolman redshift holds and a global Gibbs ensemble exists with respect to \(H_\chi\). If this is not satisfied, then \(K=0\).

\medskip
\emph{(E) Near–horizon control.} We set \(E=1\) if there is a universal near–horizon decoupling region with well posed boundary conditions such that a regulator–independent macroscopic entropy functional is obtained from a controlled path–integral principle,
\begin{equation}
S_{\rm macro}\;=\;\frac{A(\mathcal{H}^+)}{4G_N} \;+\; S_{\rm Wald}^{(\text{higher-deriv})} \;+\; S_{\rm q} 
\end{equation}
for example via an AdS\(_2\times S^{d-2}\) throat with attractor equations and the Quantum Entropy Function. Otherwise \(E=0\).

\medskip
With these definitions we may state an elementary formulation of the TSC.

\medskip
\noindent\textit{BKE criterion:} \textit{If \(B\cdot K\cdot E\neq 1\), then there is no observer independent microcanonical ensemble \(\Omega(E,\{Q_i\})\) whose degeneracy satisfies}
\begin{equation}
S_{\rm micro}\;=\;\log \Omega(E,\{Q_i\})\;=\;\frac{A(\mathcal{H}^+)}{4G_N}\;+\;O(A^0)
\end{equation}
Let us now explain what these conditions really mean and how it relates to our conjecture. The term \(B\) supplies the labels and superselection sectors against which microscopic counting is performed while the term \(K\) supplies equilibrium structure and a canonical temperature so that thermodynamic notions are global rather than patchwise and the term \(E\) supplies a regulator independent macroscopic target fixed by near horizon universality to which microscopic degeneracies can be compared. When \(B=K=E=1\), the equality \(S_{\rm micro}=S_{\rm macro}\) is realized in known stringy black holes, up to controlled subleading corrections; when any leg fails, the equality loses meaning. Note that this inequality is based on recognizing the key issues we discussed in section 3. 
\\
\\
It is important and instruct to verify \(B=K=E=1\) in some standard string constructions as well. For supersymmetric extremal black holes in \(\mathcal{N}\!\geq\!2\) compactifications, one has an asymptotic AdS region and conserved electric/magnetic charges \(\Gamma=(p^I,q_I)\), so we have \(B=1\). The near–horizon geometry takes the universal form
\begin{equation}
ds^2 \;=\; v_1\left(-r^2 dt^2 + \frac{dr^2}{r^2}\right) \;+\; v_2\, d\Omega_{d-2}^2
\end{equation}
with an extremal Killing horizon \footnote{This means that \(\kappa\!=\!0\) and hence a ground state KMS structure is there.}, so \(K=1\) in the sense of a stationary horizon. The attractor mechanism extremizes the entropy function \(E(e,p,u)=2\pi(q_ie^i-f)\), and the AdS\(_2\) path integral defines the QEF
\begin{equation}
d(\Gamma)\;=\;Z_{\rm AdS_2}[\Gamma]\;\sim\;\exp\!\Big(S_{\rm Wald}(\Gamma)+\cdots\Big)
\end{equation}
yielding \(E=1\) and so this leads to
\begin{equation}
S_{\rm micro}(\Gamma)\;=\;\log d(\Gamma)\;=\;S_{\rm macro}(\Gamma)\;=\;\frac{A}{4G_N}+ \text{corrections}
\end{equation}
For the BTZ black hole \cite{Banados:1992wn} in AdS\(_3\), Brown–Henneaux \cite{Brown:1986nw} charges at the timelike boundary give \(B=1\), the geometry is stationary with a Killing horizon so \(K=1\), and holography supplies \(E=1\) from the Cardy formula with \(c=\tfrac{3\ell}{2G_3}\). Non–extremal AdS black holes likewise have \(B=1\) and \(K=1\) although the near–horizon is not AdS\(_2\), the Euclidean saddle with contractible thermal circle and the dual CFT ensemble provide \(E=1\) in the holographic sense, again yielding equality of macro and micro entropies.
\\
\\
Now on the other hand, in FLRW cosmologies with metric \(ds^2=-dt^2+a^2(t)\gamma_{ij}dx^i dx^j\) there is no \(S^{d-2}_\infty\) on which to define ADM or brane charges invariant under time evolution and so we have \(B=0\). There is no global timelike Killing vectorand so the KMS condition cannot be implemented globally and Tolman’s relation \(T\sqrt{-\chi^2}=\text{const}\) has no meaning and so \(K=0\). There is no universal AdS\(_2\)–type throat or attractor extremization tied to a cosmological horizon and no accepted AdS\(_2\)–like boundary value problem for a QEF so any "horizon temperature" derives from quasi–local constructions and it is hence inherently slicing dependent so \(E=0\). Therefore \(B\cdot K\cdot E=0\) and the black hole style microstate equality fails to even formulate.
\\
\\
In global de Sitter one likewise has \(B=0\)  as there are no timelike boundary supporting charges. There is no globally defined timelike Killing vector, so \(K=0\) globally as within a single static patch one does have a patchwise Killing vector and a Gibbons–Hawking KMS state at \(T=H/2\pi\), but this is explicitly observer dependent and does not upgrade \(K\) to unity at the spacetime level relevant for an observer independent microstate interpretation. There is no AdS\(_2\) again, near–horizon structure nor a QEF analogue with agreed–upon measure and so \(E=0\) and so this means that the BKE criterion is again not satisfied. For anisotropic expanding Bianchi cosmologies, the situation worsens further in the case of compact homogeneous slices as it often eliminates any asymptotic sphere so \(B=0\). Further the expansion rates \(H_i\) differ and forbid a single Gibbs state which makes K equal to zero and horizon notions become directional again with no decoupled universal throat so $E=0$. This BKE criterion thus clarifies why string theoretic black hole successes are not accidents but consequences of structure.
\\
\subsection*{How the conjecture could be falsified?}
A clean way to falsify TSC is to produce a cosmological (FLRW/dS/Bianchi) setting in which the black hole pillars \((B,K,E)\) all hold so that a genuine, observer independent microstate ensemble exists and this could lead to the reproduction of the entropy area relation. To be clear, we note the following two ways one can categorically disprove this conjecture:

\paragraph{Bulk microstate geometry route:}
If one can construct in a single UV complete theory, an infinite family of smooth, horizonless cosmological microstate geometries \(\{\mathcal{M}_\alpha\}\) that are indistinguishable from a given cosmological background outside a compact region, are labeled by an observer independent invariant \(I(\alpha)\) and whose degeneracy grows as the cosmological horizon area $\sim \exp \left(\frac{A(\mathcal{H}_{\rm cos})}{4G_N}\right)$. In addition, if one can demonstrate a controlled fine grained computation that yields a unitary Page curve for the cosmological radiation with the same leading area coefficient then it would establish an observer independent \(\Omega(E,\{Q_i\})\) for a cosmological horizon and thereby would violate the generically inequivalent clause of the TSC on an open set of cosmological solutions. 
\\
\paragraph{Boundary/holography route:}
If one can exhibit a UV complete construction with a bona fide cosmology appropriate holographic dual \(\mathcal{C}\) and asymptotic charges that label superselection sectors, together with a global Gibbs/KMS structure and a controlled macroscopic entropy functional. To be clear, if one can construct a spacetime \((\mathcal{M}_{\rm cos},g)\) with a cosmological horizon \(\mathcal{H}_{\rm cos}\) and a dual Hilbert space \(\mathcal{H}_{\mathcal{C}}(E,\{Q_i\})\) such that
\begin{equation}
B=K=E=1 \quad\text{and}\quad
S_{\rm micro}(E,\{Q_i\})=\log \dim \mathcal{H}_{\mathcal{C}}(E,\{Q_i\})
= \frac{A(\mathcal{H}_{\rm cos})}{4G_N} + c_0\log A + O(A^0)
\end{equation}
with the same \(c_0\) as obtained from a regulator independent gravitational computation.
\\
\\
Either construction would realize \(B\cdot K \cdot E = 1\) in a cosmological setting and would therefore contradict the BKE criterion that is critical to the conjecture.
\\
\\
\section{Roadmap for Cosmological Horizon Entropy and Cosmological Implications}

Assuming that the TSC holds true, it tells one that a UV complete account of cosmological horizon entropy should not attempt a direct transplant of black hole methods, but instead develop cosmology native replacements for the \((B,K,E)\) pillars. We now discuss in brief what future methods can possibly focus on to achieve more quantum gravity aligned results. Note that these methods would hope to achieve entropy results which are in line with the methods for string theory as discussed before, but applicable for cosmology in general rather than black holes.  
\\
\\
One can replace asymptotic charges by cosmological invariants, which are defined on holographic screens or causal boundaries. One concrete direction is might be to use covariant phase space methods with the Kodama vector \cite{Kodama:1988yf,kod2Witten:2003mb} in spherically symmetric cosmologies to define quasilocal charges \(Q^{\rm screen}_i\) living on past or future light sheets. The goal here would be to carve the cosmological Hilbert space into well defined sectors \(\mathcal{H}_{\rm cos}=\bigoplus_{\{Q^{\rm screen}_i\}}\mathcal{H}_{\{Q^{\rm screen}_i\}}\) that survive time evolution sufficiently to support a microcanonical notion. One can also replace global equilibrium by a non-equilibrium but covariant statistical framework \cite{nonst1sohl2015deep,nonst2gallavotti2019nonequilibrium,nonst3Motta:2025xli,nonst4zwanzig2001nonequilibrium,nonst5kubo2012statistical,nonst6ruelle1999smooth}. 
\\
\\
One can also possibly replace QEF by a cosmological entropy functional, which is extremized on screens. A promising target could be a screen adapted generalized entropy
\begin{equation}
S_{\rm gen}^{\rm screen}[\Sigma] \;=\; \frac{{\rm Area}(\Sigma)}{4G_N} + S_{\rm out}(\Sigma)
\end{equation}
which is defined on codimension–two and leaves \(\Sigma\) of a holographic screen, with precise, regulator independent boundary conditions on the Wheeler–DeWitt patch bounded by the associated light sheets. From this the aim would be to construct a cosmological attractor equation
\begin{equation}
\delta S_{\rm gen}^{\rm screen}[\Sigma_\star]\;=\;0
\end{equation}
whose solution \(\Sigma_\star\) yields a universal leading term and a controlled tower of subleading corrections. Finally, one would have to make contact with observations in ways that are diagnostic of microscopic structure rather than mere curve–fitting (more on this in section 6). Non–perturbative predictions for late time correlators, null tests for BH–inspired dark energy scalings and signatures of island induced information recovery in cosmological settings can all serve to differentiate the entropy origins here of the area terms. So the important point to note is that the conjecture does not close the door on a UV description of cosmological horizon entropy but reorients it. It says now that a successful theory will likely arise not by copying \(\{B,K,E\}\) from black holes, but by discovering the cosmology–appropriate analogues.
\\
\\
After establishing the TSC so far, we now venture to its far-reaching implications on cosmology and discuss them clearly. In semiclassical GR it is common practice to assign an entropy to the cosmological event horizon in analogy with black hole entropy, as is done in the Gibbons-Hawking formulation for dS space for example ( similar idea as is taken for the Cai-Kim temperature in more general cosmological models of dark energy \cite{Cai:2005ra}). Likewise there are several models in modern cosmology explicitly build upon this analogy, deriving effective dynamics using thermodynamic arguments based on the horizon area. These include models where the Friedmann equations emerge from entropic forces, theories of emergent gravity and numerous variants of dark energy models grounded in modified entropy-area relations. One can think of many examples here like Tsallis entropy cosmology, Barrow entropy cosmology, and other generalized entropy frameworks \cite{tsaSheykhi:2018dpn,barSheykhi:2022jqq,bar2Leon:2021wyx}. 
\\
\\
But now, if string theory suggests that a proper quantum gravity description of cosmological horizon entropy is different from its Black hole analogue, then the validity of these thermodynamic derivations becomes questionable. The use of entropy-area formulas and associated thermodynamic reasoning in cosmology may amount to an extrapolation beyond the domain of validity of the semiclassical framework. This perspective motivates a deeper reconsideration of the fundamental thermodynamic assumptions underlying several cosmological paradigms. For instance, this could propose revisions for the entropic gravity proposal, where we see that gravity itself is viewed as an emergent phenomenon arising from changes in entropy associated with holographic screens \cite{Verlinde:2010hp} and it assumes that entropy-area relations hold in a cosmological setting. Similarly we see that emergent spacetime paradigms (like the Padmanbhan framework or other such approaches \cite{Padmanabhan:2009vy}) that utilize thermodynamic identities to recover the Einstein equations often rely implicitly on the assumption that cosmological horizons can be treated thermodynamically in a fashion analogous to black holes. If UV complete gravity prohibits this analogy, these derivations may not merely be approximations but may be qualitatively incorrect in their foundations. Even more striking are the implications for models of dark energy inspired by holography as HDE models \cite{hderWang:2016och} posit that the energy density of dark energy is related to the largest distance scale in the universe, typically the future event horizon, constrained by an entropy bound and so, this goes to questioning the consistency of the entire framework of HDE models within a UV complete theory of quantum gravity. It would, largely mean, that all such approaches may lie in the Swampland \cite{swa1Vafa:2005ui,swa2Ooguri:2006in,swa3Palti:2019pca,swa4Obied:2018sgi,swa5Ooguri:2018wrx,swa6Garg:2018reu,swa7Agrawal:2018own,swa8Brennan:2017rbf,swa9Ooguri:2016pdq,swa10Lust:2019zwm,swa11Akrami:2018ylq,swa12vanBeest:2021lhn}.
\\
\\
It is also important here to appreciate that this line of reasoning does not merely restrict specific models, but rather casts doubt on a broad methodological approach. To put it more specifically, it is geared towards questioning the practice of extrapolating black hole thermodynamic insights to cosmological horizons. If the underlying microphysics is absent or categorically different then such an extrapolation may misrepresent the true behavior of gravitational dynamics at the cosmological scale. Instead, a consistent approach to cosmological thermodynamics in quantum gravity may require a completely different conceptual framework, as discussed previously in this section, and it is likely one which is not directly modeled on black hole physics.
\\
\\
\section{Observational Test for Bekenstein–Hawking Cosmological Horizon Entropy}
Now that we have established an argument about a quantum gravity description possibly requiring a rethink of Bekenstein–Hawking relation for cosmological horizons, we can go a step further. We can ask a critical question which is; could there be an observational way to check the applicability of the relation? If one could establish using observational data that the relation holds good or otherwise, then that would directly either prove or falsify the conjecture. If the conjecture is proved, then it shows conclusively that quantum gravity requires some rethink of thermodynamics for different horizons. If it is proven false, then it shows that while the Bekenstein–Hawking relation holds good, one has to think of new string theory methods to reproduce that formula from microscopic principles being applicable to an expanding cosmology.
\\
\\
We begin by assuming that the Bekenstein–Hawking prescription applied to a cosmological event horizon is fine and so it were to hold, the entropy as a function of redshift $z$ is
\begin{equation}
S(z)\;=\;\frac{\pi k_B c^3}{G\hbar}\;R_{\rm eh}^2(z)
\end{equation}
with the event-horizon proper radius
\begin{equation}
R_{\rm eh}(z)\;=\;\frac{c}{1+z}\;\int_{z}^{-1}\frac{dz'}{H(z')}
\end{equation}
where the upper limit $z'=-1$ denotes the infinite future ($a\to\infty$). In general this would render the integral mildly model dependent but in the case of an accelerating universe at late times one has the familiar approximation $R_{\rm eh}(z)\approx c/H(z)$, yielding a simple form
\begin{equation}
S(z)\;\approx\;\frac{\pi k_B c^5}{G\hbar}\;\frac{1}{H^2(z)}
\label{eq:SBH-H}
\end{equation}
entirely in terms of the Hubble rate $H(z)$ that is independently measurable from BAO, supernovae, cosmic chronometers and standard sirens.
\\
\\
The central observational difficulty is that one never directly measures a “horizon entropy.” What one measures are temperature or intensity maps, correlation functions, distances, and the growth of structure and so our strategy here is to therefore  construct data native entropy proxies $S_{\rm LHS}(z)$ that can be computed directly from maps in observable space (angles and frequencies) without inserting $H(z)$ at any stage of their definition. We can then test whether their redshift evolution follows the Hawking scaling implied by \eqref{eq:SBH-H} and concretely, we will compare $S_{\rm LHS}(z)$ to the right-hand side $S_{\rm BH}(z)\propto H^{-2}(z)$ through an exponent test $S_{\rm LHS}(z)\propto H^{-\beta}(z)$.
\\
\\
Consider now a 21-cm tomographic data cube $T(\hat{\boldsymbol n},\nu)$ or a cleaned CMB map at an effective redshift slice $z$. We work entirely now in terms of observables, with those being the angular direction $\hat{\boldsymbol n}$ is measured on the sky, and the frequency $\nu$ which tags the slice through the rest frequency of the line, like $1+z=\nu_{21}/\nu$ for 21-cm. Note interestingly here that no cosmology is invoked here to assign a slice to $z$. In a flat sky approximation, Fourier modes are labeled by transverse multipole $\ell$ and line-of-sight delay $\eta$ (which is in general considered the Fourier dual to frequency). For each resolved mode we define a per mode Gaussian information content
\begin{equation}
I_{\boldsymbol k}(z)\;=\;\frac{1}{2}\log\!\Big(1+\mathrm{SNR}^2_{\boldsymbol k}(z)\Big),\qquad 
\mathrm{SNR}^2_{\boldsymbol k}(z)\;=\;\frac{P_s(\boldsymbol k,z)}{P_n(\boldsymbol k,z)}
\end{equation}
where $P_s$ and $P_n$ are, respectively, the measured signal and noise power in that mode after instrument calibration, beam deconvolution $B_\ell$, bandpass and foreground treatment. Summing over all modes in the survey window $\mathcal{K}(z)$ gives an empirical information capacity
\begin{equation}
\mathcal{I}(z)\;=\;\sum_{\boldsymbol k\in\mathcal{K}(z)} \frac{1}{2}\log\!\Big(1+\mathrm{SNR}^2_{\boldsymbol k}(z)\Big)
\;\approx\;\frac{V_{\rm obs}(z)}{2(2\pi)^3}\int_{\mathcal{K}(z)} d^3k\;\log\!\Big(1+\mathrm{SNR}^2_{\boldsymbol k}(z)\Big)
\end{equation}
from which we define the information entropy proxy
\begin{equation}
S_{\rm info}(z)\;\equiv\;k_B\,\mathcal{I}(z)
\end{equation}
By construction, we see that $S_{\rm info}(z)$ is entirely data-defined. It is so as it depends on the number and quality of statistically independent modes actually resolved in the slice and does not require any conversion to comoving units or the insertion of $H(z)$. If the ultimate causal ceiling on independent information in a slice scales effectively like a horizon area, then the redshift evolution of $S_{\rm info}(z)$ should obey a power law in the instantaneous causal scale which is
\begin{equation}
S_{\rm info}(z)\;\propto\;R_{\rm eh}^{\,\beta}(z)\qquad\Longleftrightarrow\qquad S_{\rm info}(z)\;\propto\;H(z)^{-\beta}
\end{equation}
with Hawking’s area law corresponding to $\beta=2$ in this scheme.
\\
\\
Another entropy proxy can be based on the empirical Shannon entropy \cite{shannon1948mathematical} of a cleaned map slice. This can be done after whitening the map to remove known instrumental correlations and fixing a canonical serialization, and we can then define
\begin{equation}
S_{\rm pix}(z)\;=\;-k_B\sum_{b} p_b(z)\,\log p_b(z),\qquad 
S_{\rm zip}(z)\;\equiv\;k_B\,L_{\rm zip}(z)
\end{equation}
where $p_b(z)$ is the observed histogram of pixel intensities in bin $b$, and $L_{\rm zip}(z)$ is the length in bits of a losslessly compressed file of the map under a fixed compressor. As nonparametric complexity measures, $S_{\rm pix}(z)$ and $S_{\rm zip}(z)$ capture how many degrees of freedom are needed to describe the slice and they are measured directly from the data in angle-frequency space and again require no knowledge of $H(z)$ to define and so if cosmological information is limited by an area like causal screen then one would expect
\begin{equation}
S_{\rm pix/zip}(z)\;\propto\;H(z)^{-\beta}
\end{equation}
with $\beta\approx 2$; a volume-limited ceiling would instead lean toward $\beta\approx 3$ in idealized limits, providing a discriminant.
\\
\\
Having defined left hand side (LHS) entropy proxies purely from data, we now specify the right hand side (RHS) of \eqref{eq:SBH-H} that encodes the Hawking prediction. Using independent determinations of $H(z)$ (BAO scale measurements, cosmic chronometers, standard sirens etc.) we know that
\begin{equation}
S_{\rm RHS}(z)\;=\;\frac{\pi k_B c^5}{G\hbar\,H^2(z)}
\end{equation}
We then pose a scaling test that is blind to overall normalization uncertainties by fitting an exponent $\beta$ through
\begin{equation}
S_{\rm LHS}(z)\;=\;A\,H(z)^{-\beta}\,,
\qquad\Longleftrightarrow\qquad
\log S_{\rm LHS}(z)\;=\;\log A\;-\;\beta\,\log H(z)
\end{equation}
over multiple redshift bins with full covariance (from end to end simulations), marginalizing known instrumental systematics and selection functions. A measurement $\beta=2\pm\sigma_\beta$ would support the Hawking area scaling for cosmological horizons while a statistically significant deviation would indicate either modified area–entropy relations or the breakdown of the area-law interpretation for cosmological horizons, thereby informing the conjecture. In either case, it would signal a departure from black hole thermodynamics for an expanding cosmology.
\\
\\
It is important to emphasize why this comparison is not circular as the LHS quantities $S_{\rm info}(z)$ and $S_{\rm pix/zip}(z)$ are constructed entirely in observable space (angles and frequencies), with the redshift of each slice determined by the rest frequency of the observed line. None of their definitions requires converting to comoving distances, volumes, or horizon sizes, nor do they invoke $H(z)$. The only use of $H(z)$ is on the RHS, where it enters the Bekenstein–Hawking prediction being tested. So, the LHS furnishes an empirically determined entropy-versus-redshift curve and the RHS imposes a scaling law $H^{-2}(z)$ against which this curve is compared.
\\
\\
From a practical standpoint, current and near future facilities already permit these kind of implementations. During the Epoch of Reionization, HERA \cite{hera1DeBoer:2016tnn,hera2HERA:2021noe} and SKA-Low Phase 1 \cite{ska1Weltman:2018zrl,ska2Maartens:2015mra,ska3Santos:2015gra} can deliver $S_{\rm info}(z)$ and $S_{\rm topo}(z)$ in a few broad redshift bins using foreground-avoidance windows. This would thus be providing initial constraints on $\beta$ albeit with sizable uncertainties. With next-generation tomography (full SKA-Low and mid-band intensity mapping with long baselines and deep integrations), one expects access to billions of independent modes across $6\lesssim z\lesssim 30$. This would end up in  enabling precise measurements of $\mathcal{I}(z)$ and topology and driving $\sigma_\beta$ down to the regime where $\beta=2$ can be decisively confirmed or refuted. Cross checks with small scale CMB temperature/polarization maps furnish an anchor at $z\simeq 1100$ \cite{Planck:2018vyg}, while line-intensity mapping in the range $2\lesssim z\lesssim 6$ bridges to late time structure \cite{Kovetz:2019uss}. The unifying idea here is that a genuine observational test of cosmological horizon entropy must extract an $S_{\rm LHS}(z)$ whose redshift dependence is empirically measurable from data alone and only then is any comparison to $S_{\rm BH}(z)\propto H^{-2}(z)$ a sharp, model differentiating test of the Bekenstein–Hawking relation in an expanding cosmology.
\\
\\
\section{Conclusions}
In this work we began by surveying the principal string theoretic routes to black hole entropy. These included D–brane microstate counting and AdS/CFT Cardy asymptotics, the quantum entropy function with attractors and Wald corrections, OSV/topological string relations, the string–black hole correspondence, and island based fine grained entropy formalism. We then noted the common structural ingredients that make these successes possible and argued how these structures fail for the case of an expanding cosmology. This structural mismatch underlies the failure of direct transplants of black hole logic to cosmological horizons and motivates our \textit{Thermodynamic Split Conjecture} (TSC), which is the statement that in a UV complete quantum gravity, the microscopic and thermodynamic structures of black–hole and cosmological horizons are generically inequivalent. We then noted how these ingredients can be encoded in the triad \((B,K,E)\) and led us to proposing the BKE criterion.
\\
\\
After this, we then made the case for two possible ways to falsify the conjecture by theoretical proofs(a cosmology-grade holography with charge-labeled ensembles reproducing the area law or a bulk construction of cosmological microstate geometries with area-growth degeneracy) and noted the implications of all this on cosmology as a whole. We then proposed an observational program that compares data native entropy proxies from tomographic maps to the Bekenstein–Hawking scaling \(S\propto H^{-2}(z)\). Either outcome that may come from this observational avenue would be informative, as confirmation of the Hawking scaling across redshift would press for new string-theoretic microphysics tailored to expanding universes, while a departure would vindicate the TSC and reorient theory toward cosmology-native replacements for \((B,K,E)\). Near term experiments can deliver strong constraints on this and next-generation tomography should render the exponent test decisive. Formal developments in lower-dimensional gravities, static-patch/dS holography etc. in cosmological settings are natural next steps prompted by the conjecture. Overall, this work also shows an exciting amalgamation of how ideas from black hole physics, thermodynamics, string theory, observational and theoretical cosmology and information theory, can all come together to guide us towards new avenues of understanding.
\\
\\
\section*{Acknowledgements}
The author is supported in part by the Vanderbilt Discovery Doctoral Fellowship. The author would also like to thank Robert Scherrer, Paul Steinhardt, Abraham Loeb, Samir Mathur and Alexandru Lupsasca for helpful discussions.

\bibliography{references}
\bibliographystyle{unsrt}
\end{document}